\documentclass[traditabstract]{aa}
\usepackage{graphicx,xcolor,ulem}
\usepackage{lscape}
\usepackage{units,natbib}
\usepackage{txfonts}
\begin{document}
\newcommand{\com}[1]{\textbf{\color{blue}{#1}}}
\newcommand{\edit}[2]{\textbf{\color{red} {#1}}}
\newcommand{\so}[1]{\textbf{\color{red} \sout{#1}}}
\newcommand{\vs}{{\it vs\ }}
\newcommand{\kms}{km~s$^{-1}$}
\newcommand{\Msun}{M$_{\odot}$}
\newcommand{\msun}{M_{\odot}}
\newcommand{\myr}{M$_{\odot}$\,yr$^{-1}$}
\newcommand{\Teff}{$T_{\rm eff}$}
\newcommand{\FeH}{[Fe/H]}

\title {The formation of long-period eccentric binaries with a helium white dwarf}

   \author{L. Siess
          \and P.~J. Davis
          \and A. Jorissen
  }

  \institute{Institut d'Astronomie et d'Astrophysique, Universit\'e Libre
    de Bruxelles, ULB, CP. 226, Boulevard du Triomphe, 1050 Brussels, Belgium \\
    \email{siess@astro.ulb.ac.be} }

\date{Received ...; accepted ...}

\abstract {The recent discovery of long-period eccentric binaries hosting
  a He-WD or a sdB star has been challenging binary-star modelling. Based
  on accurate determinations of the stellar and orbital parameters for IP
  Eri, a K0 + He-WD system, we propose an evolutionary path that is able to
  explain the observational properties of this system and, in particular,
  to account for its high eccentricity (0.25). Our scenario invokes an
  enhanced-wind mass loss on the first red giant branch (RGB) in order to avoid mass    
  transfer by Roche-lobe
  overflow, where tides systematically circularize the orbit. We explore
  how the evolution of the orbital parameters depends on the initial
  conditions and show that eccentricity can be preserved and even increased
  if the initial separation is large enough. The low spin velocity of the
  K0 giant implies that accretion of angular momentum from a (tidally-enhanced) RGB wind should not be
  efficient. }

\keywords{binaries:general -- white dwarfs -- stars:evolution
  -- stars:individual:IP Eri}


\maketitle

%

\section{Introduction}

The recent discovery of several long-period binaries ($P\sim10^3$~d)
hosting a sdB star
\citep{Ostensen2011,Ostensen2012,Vos2012,Deca2012,Barlow2012,Barlow2013} or a
He white dwarf (WD) like HR 1608 = 63~Eri \citep{Landsman1993,Vennes1998}
or IP~Eri studied here \citep{Merle2014} have challenged our
understanding of their formation. These two classes of stars are closely
related, since sdB stars consist of He-burning cores surrounded by
extremely thin H envelopes \citep{Heber1986}, while He WDs are degenerate,
nuclearly extinct He cores.

The formation of sdB stars and He WDs require that the progenitor star
loses its envelope as it ascends the red giant branch (RGB). The most
likely scenario, if not the only one for the He WDs, requires a binary
companion. Among the 5 different evolutionary channels proposed by
\cite{Han2002,Han2003} for the formation of sdBs, the only one that leads
to long-period systems involves stable mass transfer on the RGB.  To avoid
dynamical Roche lobe overflow (RLOF) and subsequent common envelope (CE)
evolution, the initial system mass ratio must be less than some critical
value of the order of $\sim 1.2-1.5$
\citep[e.g.][]{Webbink88,Soberman97}. This scenario is likely to produce
circular systems with periods $P \la 500$~d \citep{Podsi2008} that may be
too short to account for the thousand days period of the previously
mentioned objects. An alternative scenario has emerged from the binary
population-synthesis models of \citet{Nelemans2010}. Using the
$\gamma$-prescription for the CE efficiency \citep{Nelemans2005}, which is
based on the angular momentum balance rather than the energy balance, the
authors showed that CE channels can produce systems with main-sequence
companions and periods on the order of years but likely to be circular.

More recently, \citet{Clausen2011} proposed a different evolutionary path
leading to eccentric, long-period sdB + main-sequence (MS) binary systems,
starting from hierarchical triple systems whose inner binaries merge and
form sdBs, while the outer MS star had no part in the sdB formation. Thus,
unlike stable RLOF- and CE-produced systems, which should have nearly
circular orbits, no limitations exist on the eccentricities of these
binaries other than a requirement that the periastron separation of the
outer binary not be too small. This results in eccentric systems with final
orbital periods on the order of 1000~d. The application of this scenario to
the formation of a long-period eccentric binary involving a He WD is more
problematic, because it requires the merger product of the inner binary to be a He WD, i.e. with a mass below $\sim 0.45 M_\odot$.

Based on state-of-the-art binary-evolution calculations done with BINSTAR, we
present a consistent evolutionary channel to explain the properties of
systems like IP~Eri involving a He WD in a long-period eccentric orbit. The
paper is organized as follows: in the next two sections we summarize the
main physical ingredients of BINSTAR and properties of IP Eri. Then in
Sects.~\ref{Sect:rlof} and \ref{Sect:crap}, we present the results of
our calculations for the RLOF and tidally-enhanced wind models
and conclude in Sect.~\ref{Sect:conclusion}.

\section{The binary evolution code \textsc{BINSTAR}}

The BINSTAR code \citep{Siess2013,Davis2013,Deschamps2013} used in this
work is based on the stellar evolution code STAREVOL \citep{Siess2006} and
is specifically designed to study low- and intermediate-mass binaries.  The
evolution of the orbital parameters (semi-major axis and eccentricity) is
calculated simultaneously along side the internal structure and rotation of
the two stellar components.

The evolution of the orbital parameters is governed by the conservation of
the system's angular momentum (hereafter AM, $J_{\Sigma}$), according to
the relation
\begin{equation}
 \dot{J}_{\Sigma} = \dot{J}_{d} + \dot{J}_{g} + \dot{J}_{\mathrm{orb}} 
\label{eq:Jorb}
\end{equation}
where $\dot{J}_{\mathrm{orb}},\ \dot{J}_{d}$ and $\dot{J}_{g}$ are the
torques applied to the orbit, donor and gainer star, respectively. The
stellar torques that act on the stellar spins account for the action of
tides (synchronization) and change in AM due to the accretion or loss of
mass. Note that in all the simulations, we consider solid-body rotation and
assume the initial stars to be (pseudo-)synchronized. The evolution of the
semi-major axis ($a$) is given by
\begin{equation}
  \frac{\dot{a}}{a}=2\frac{\dot{J}_{\mathrm{orb}}}{J_{\mathrm{orb}}}-
  2\left(\frac{\dot{M}_{d}}{M_{d}}+\frac{\dot{M}_{g}}{M_{g}}\right)+ 
  \frac{\dot{M}_{d}+\dot{M}_{g}}{M_{d}+M_{g}}+\frac{2e\dot{e}}{1-e^{2}} 
\label{eq:adot}
\end{equation}
where $\dot{J}_{\mathrm{orb}} $ is evaluated from Eq.~(\ref{eq:Jorb}) and
$\dot{M}_{i=d,g}$ corresponds to the net\footnote{This includes
  contributions from mass accretion/loss  via RLOF and/or wind.} mass
change rate of star $i$.  RLOF mass transfer rates are computed according
to the \cite{Kolb1990} prescription and the \cite{Eggleton1983}
formulation for the Roche lobe radius ($R_{L1}$) is used.

Changes in the eccentricity ($e$) due to the tidal interaction of each star
($\dot{e}_{\mathrm{tide},i}$) are calculated from \cite{Zahn77,Zahn89}
\cite[for details, see][hereafter Paper~I]{Siess2013}. However in an
eccentric orbit, the specific AM of the ejected material depends on the
star's position along its orbit and is potentially able to generate some
eccentricity. Following \cite{Soker2000}
\begin{equation}
  \dot{e}_{\mathrm{winds}}(\nu)=\frac{\bigl|\dot{M}_{g}^{\mathrm{wind}}+
  \dot{M}_{d}^{\mathrm{wind}}\bigr|}{(M_d+M_g)}\,(e+\cos\nu) 
\label{eq:edot_wind}
\end{equation}
where $\nu$ is the true anomaly and
$\dot{M}_{i}^{\mathrm{wind}}$ the wind mass loss rate from star
$i$, assumed in these calculations to follow the
\cite{Reimers1975} prescription. The total rate of change of the
eccentricity is given by
\begin{equation}
\dot{e} =
\dot{e}_{\mathrm{winds}}+\dot{e}_{\mathrm{tide},d}+\dot{e}_{\mathrm{tide},g}\ .
\label{eq:edot}
\end{equation}
The specific AM of the wind material ejected from the system is assumed to
be equal to the specific orbital AM of the star at its position on the
orbit, i.e.
\begin{equation}
\dot{J}_\Sigma = \sum_{i=g,d}\dot M^{\mathrm{wind}}_i a_i^2\omega =
\Biggl(\frac{\dot{M}^{\mathrm{wind}}_d}{q}+q \dot{M}^{\mathrm{wind}}_g\Biggr)\, j_\mathrm{orb}
\label{eq:jdot}
\end{equation}
where $\omega$ is the orbital angular velocity, $a_i$ the distance of star
$i$ to the center of mass of the system ($a_1+a_2=a$), $q= M_d/M_g$ the
mass ratio and $j_\mathrm{orb} = J_\mathrm{orb}/(M_d+M_g)$
the specific orbital AM.

When the mass exchange rate depends on the orbital phase, as is the case in
an eccentric orbit for the RLOF mass transfer rate and tidally enhanced
stellar winds (see Sect.\ref{Sect:eccentric}), averaged quantities must be
used in order to follow the secular evolution of the system over millions
of orbits. To achieve this goal, we use a Gaussian quadrature integration
scheme which involves the calculation of the mass transfer properties at
very specific locations (i.e. at given true anomalies) along the
orbit. Once the instantaneous Roche radius is known, at that orbital phase
one can calculate the tidal torques ($\dot{e}_{\mathrm{tide},i}$) and the
mass transfer rates (either due to RLOF or tidally enhanced wind) from
which the ``local'' values of $\dot{e}_{\mathrm{winds}}$ and $\dot J_{i}$
are derived. The mean quantities $\langle \dot{M}_i \rangle$, $\langle
\dot{e}_{\mathrm{winds}} \rangle$ and $\langle \dot{a}/a \rangle$ are
then simply calculated by summing the weighted contribution of
each variable at the specified points. 
Technically, this resumes to calculating for $X=\{\dot{M}, \dot{e}, \dot{a}\}$
\begin{eqnarray*}
\langle X \rangle & = & \frac{1}{P}\int_0^P X(t) \,dt= \frac{(1-e^2)^{3/2}}{\pi}\int_0^{2\pi}
\frac{X(\nu)}{(1+e\cos\nu)^2}\, d\nu \\
 & \approx & (1-e^2)^{3/2} \sum_{i=1}^N w_i\, 
 \frac{X(\nu_i)}{(1+e\cos\nu_i)^2}
\end{eqnarray*}
where $\nu_i$ and $w_i$ are tabulated coefficients.\footnote{see
  e.g. http://en.wikipedia.org/wiki/Gaussian\_quadrature}

\section{Observational and evolutionary constraints}
\label{sect:obs}

\citet{Merle2014} derive the first orbital solution for IP~Eri
(= HD 18131 = EUVE J0254-053), a K0IV + DA WD system with $P =
1071.00\pm0.07$~d and $e = 0.25\pm 0.01$. An analysis of the WD atmospheric
lines by \citet{Burleigh1997} and \citet{Vennes1998} yielded an effective
temperature of $\sim 30\,000$~K and gravity $\log g \sim 7.5$. These values,
combined with structural models and a revised Hipparcos distance estimate of $\sim
100 {+26 \atop -7}$~pc, leads to a mass close to $0.4\pm 0.03\;\msun$ for the hot
companion, implying that the white dwarf is made of He and not of
carbon-oxygen. The stellar metallicity of the K0 giant is close to solar
with [Fe/H]~$\sim 0.1$, with an effective temperature $T_\mathrm{eff} =
4960\pm 100$~K and $\log g = 3.3 \pm 0.3$. These
parameters indicate that the star is located at the base of the RGB in the
Hertzsprung-Russell diagram (HRD) and its initial mass ranges between
$1.2-1.3 \la M_\mathrm{giant}/\msun \la 3$.  The low system mass function
($f = 0.0036$~\Msun) only sets an upper limit on $M_\mathrm{giant} <
4.26$~\Msun. A detailed chemical analysis of the giant reveals no s-process
enhancement, implying that the WD progenitor avoided the asymptotic giant
branch (AGB) phase, consistent with its He composition. No sign of fast
rotation was detected in the giant with $v\sin i< 5$~\kms.

Given the mass of the He WD, its progenitor must have been less massive
than 3~\Msun{} initially, because stars of higher masses leave the main
sequence with a H-depleted core more massive than the derived WD mass.  We
also inferred from single-star calculations performed with STAREVOL, that
the age of the He WD on its cooling track is $\approx 10^7$yr.

Evolutionary timescales also impose some constraints on the mass of the WD
progenitor. If we consider the system to be $\sim 5$~Gyr old, compatible
with its solar-like composition, this imposes the mass-losing star to be at
least $\sim 1.2-1.3$~\Msun{}, so that it can reach the RGB and start losing mass
in that time interval.

Furthermore, the initial mass ratio must be close to unity so that, by the
time the He WD forms and reaches its observed position in the HRD, the
companion star has evolved substantially off the main sequence to comply
with the observed values.

\section{Stable RLOF channel}
\label{Sect:rlof}
\subsection{Circular systems}
\label{Sect:circular}

We start by investigating the stable RLOF channel, first for circular
systems (Sect.~\ref{Sect:circular}) and then for eccentric ones
(Sect.~\ref{Sect:eccentric}). The key point in this approach is to start
with a binary having a mass ratio just above unity, so that when RLOF
starts, $q$ rapidly drops below unity ensuring stable mass transfer. The
initial orbital period is chosen in such a way that the donor star fills its Roche lobe on
the RGB. A set of evolutionary tracks has been computed for systems with
initial masses $M_1 = 1.2$ and $M_2 = 1.0$~\Msun\ and initial periods of
30, 100, 200 and 365~d.

\begin{figure}
\includegraphics[width=9cm]{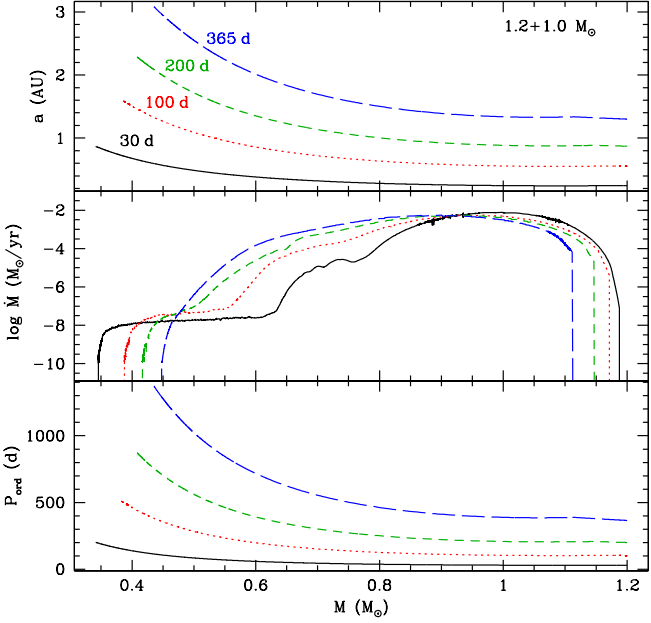}
\caption[]{\label{Fig:rlof} Evolution of the semi-major axis (top), mass-transfer rate (mid) and period (bottom, in days) for the
  circular RLOF systems as a function of the donor's mass. The curves
  correspond to different initial orbital periods: 30 (solid,
  black), 100 (red, dotted), 200 (green, short-dashed) and 365~days
  (blue, long-dashed). The initial masses are 1.2 + 1.0~\Msun.}
\end{figure}

As shown in Fig.~\ref{Fig:rlof}, the final orbital periods range from 200 to
1400~d, covering the observed value.  With increasing initial periods, mass
transfer starts at higher luminosities corresponding to larger core
masses. As a consequence, the mass of He WD increases from 0.34~\Msun{} to
0.44~\Msun. In all simulations, the mass-transfer rate peaks around
$10^{-2}$~\myr, decreases once the mass ratio has been reversed and
then stabilizes around $10^{-8}$~\myr. We note that the phase of rapid mass
transfer ($\dot M > 10^{-5}$~\myr) occurs over a larger mass range if RLOF
starts later in the evolution. The reason for this behavior is because, as
the star climbs the RGB, the energy demand from the H-burning shell (HBS)
increases.  To compensate for the higher radiative losses from the surface,
H is burnt at a higher rate and the core growth rate ($\dot
M_\mathrm{HBS}$) increases.  Because the expansion of the star is driven by
the HBS, a higher $\dot M_\mathrm{HBS}$ leads to a larger overfilling factor
($R/R_{L1}$) and hence mass-transfer rate. When all the envelope is
removed, the star leaves the RGB and moves to the blue as the hot He core
is progressively exposed. Our simulations can reproduce the observed period
and WD mass of IP Eri provided the system had an initial period of $\approx 300$~days.

The evolution of the secondary component is not very sensitive to the
initial period. It lands on the main sequence with a mass on the order of
1.6 -- 1.7~\Msun\ when a conservative evolution is considered. As described
at the end of Sect.~\ref{Sect:crap}, some fine tuning is required to
match the position in the HR diagram for both stars. It is also worth
noticing that in this process, the accretion of AM spins up the gainer close
to its breakup velocity which is incompatible with observations.

\subsection{Eccentric systems}
\label{Sect:eccentric}

If we now consider a system with an initial eccentricity $e=0.3$ and
re-perform the previous calculations, we find that in all cases, the
orbit has circularized by the time RLOF starts. The reason is that mass
transfer begins when the star has already evolved along the giant branch
and possesses a deep convective envelope. Convection is the most efficient
mechanism for dissipating the kinetic energy of the tidally-induced large-scale flows \cite[e.g.][]{zahn2008}. According to \cite{zahn78}, the
circularization timescale can be expressed as
\begin{equation}
  \frac{1}{\tau_\mathrm{circ}}  = \frac{1}{84 \tilde{q}(1+\tilde{q})\,k_{2}}
  \biggl(\frac{MR^{2}}{L}\biggr)^{1/3} \biggl(\frac{R}{a}\biggr)^{8} 
  \approx   \frac{2}{\tilde{q}(1+\tilde{q})}\biggl(\frac{R}{a}\biggr)^{8}
  \mathrm{yr} 
\label{eq:tau_circ} 
\end{equation} 
where $k_2$ is the apsidal motion constant and $\tilde{q}=M_i/M_{3-i}$, the
other quantities having their usual meanings. Inserting the Roche lobe
radius ($R_{L1}$) formula of \cite{Pac71} in Eq.~(\ref{eq:tau_circ}) and
assuming $q\approx 1$ as required by our scenario, we find that
$\tau_\mathrm{circ} \approx 4600 (\nicefrac{R_{L1}}{R})^8$~yr which is
extremely short in comparison to the evolutionary (Kelvin-Helmholtz)
timescale along the RGB. Therefore, as the star expands and progressively
fills its Roche lobe, the orbit circularizes before RLOF had even
started. So, within this paradigm, it is impossible to prevent the
circularization of the orbit, at least for the low- and intermediate-mass
stars that we consider.

\section{The enhanced wind scenario}
\label{Sect:crap}

The only solution to form a He WD and preserve some eccentricity is to
remove mass from the giant while keeping it well inside its Roche radius so
that the tidal effects remain weak. One possibility to meet these
requirements is to boost the wind mass-loss rate prior to RLOF as proposed
by \cite{CRAP88}. In their model, the authors advocate that tidal
interactions and/or magnetic activity are responsible for the stellar wind
enhancement and assume that the multiplying factor has the same dependence
on the radius and separation as the tidal torque applied onto that
star. This lead \cite{CRAP88} to propose the following expression for the
mass-loss rate
\begin{equation}
  \dot{M}^\mathrm{wind}_i = \dot{M}_i^\mathrm{Reimers}\times \Biggl\{1+B_\mathrm{wind}\times
  \min\biggl[\biggl(\frac{R}{R_{L1}}\biggr)^6,\frac{1}{2^6}\biggl]\Biggl\}\ .
\label{eq:mloss}
\end{equation}
where $\dot{M}^{\mathrm{Reimers}}$ is the Reimers' mass-loss rate and the
constant $B_{\mathrm{wind}}=10^4$ was found to match the properties of Z
Her, a RS CVn system with a mass ratio below unity.

\begin{figure}
\includegraphics[width=9cm]{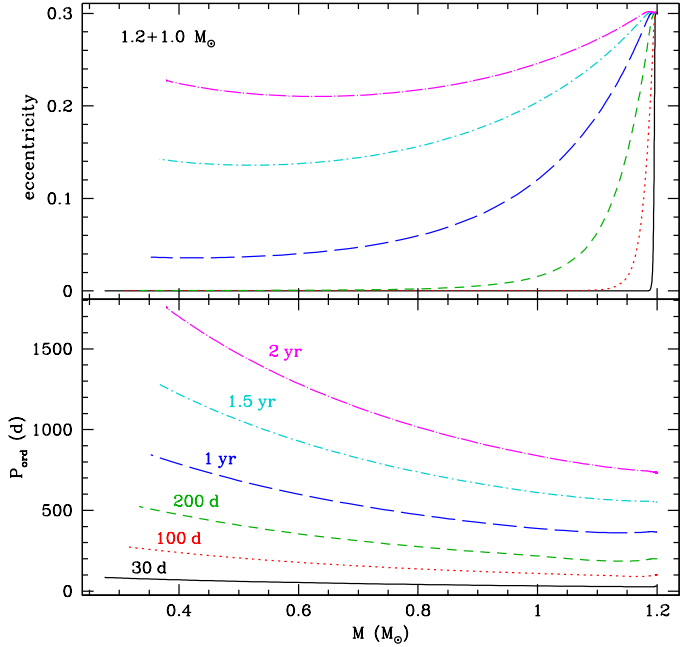}
\caption[]{\label{Fig:crap} Evolution of the eccentricity (top) and orbital
  period (bottom, in days) for the enhanced-wind models as a function of
  the donor's mass. The curves correspond to different initial orbital
  periods: 30 (solid, black), 100 (red, dotted), 200 (green,
  short-dashed) days, 1 (blue, long-dashed), 1.5 (cyan, dot-short dashed)
  and 2 years (magenta, dot-long dashed). The initial stellar masses are
  1.2 + 1.0~\Msun.}
\end{figure}

The results of our new simulations including the tidally enhanced wind are
depicted in Fig.~\ref{Fig:crap} (we use the same initial conditions as in
the previous section). Several points draw our attention: first, this
mechanism leads to smaller He WD masses because, for a given initial
period, the envelope is removed faster (to prevent the radius from
approaching $R_{L1}$ too closely) which in turn leaves less time for the 
H-burning shell to advance outward. Second, the higher amount of mass lost by
the system reduces the final separation compared to the previous
(conservative) RLOF evolution. For example, in the RLOF calculation, the one-year initial-period system leads to the formation of a
1400~d period system with a 0.43~\Msun{} He WD. For the enhanced wind
prescription, we obtain a 0.36~\Msun\ He WD binary in a 850~d orbital
period. This implies that in order to reproduce the observed period of IP
Eri, a larger initial separation must be selected.

However, the most interesting feature is the preservation and even increase
of the eccentricity in the long-period systems. We remark that, if the
initial separation is less than $\sim 200$~d for our 1.2+1.0~\Msun{} system, the
mass-loss rate enhancement occurs too early, when
$\dot{M}_i^{\mathrm{wind}}$ is relatively low. As a consequence, the
eccentricity pumping term ($\dot{e}_\mathrm{winds}$) in Eq.(\ref{eq:edot})
remains small, $|\dot{e}_\mathrm{tides}|$ dominates and the eccentricity
globally decreases. Thus, there exists a critical initial period above
which $\dot{e}_\mathrm{winds} > |\dot{e}_\mathrm{tides}|$ and eccentricity
can be preserved but its analytical derivation is not straightforward
because of the dependence of the mass loss rate on the eccentricity (via
the determination of the Roche radius) and the simplistic approach of
\cite{Soker2000}, which assumes a constant mass-loss rate independent of
the orbital phase, cannot be reiterated here.

\begin{figure}
\includegraphics[width=9cm]{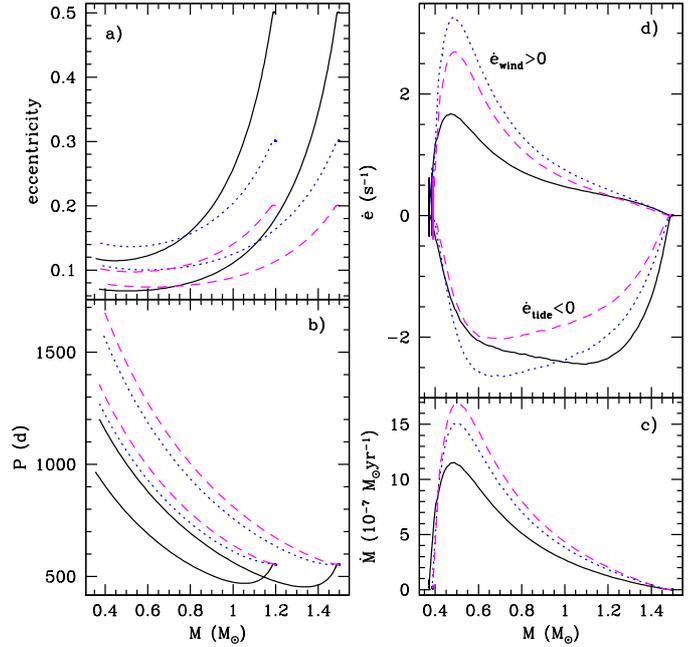}
\caption[]{\label{Fig:IP_phys} Evolution of the eccentricity (a), orbital
  period (b), wind mass loss rate (c) and $\dot e$ contributions (d) as a
  function of the donor's mass under various physical configurations. The
  initial period is the same for all simulations and equal to 550 days. The
  three curves in panel (a) and (b) starting at $M=1.5$~\Msun{} show the
  evolution of a 1.5+1.0~\Msun{}  system for initial eccentricities equal to 0.5
  (black, solid), 0.3 (blue, dotted) and 0.2 (red, short-dashed). The
  evolution of a 1.2+1.0~\Msun{}  system with same initial eccentricities is also
  shown with the same line coding but different colors. In the right
  panels, only the properties of the 1.5+1.0~\Msun{}  system are shown. The positive
  $\dot{e}$ corresponds to the pumping term, the negative part to the tidal
  one.}
\end{figure}

However, the final eccentricity still remains lower than the observed
value. To improve the situation, we explored the parameter space, first
varying the initial mass ratio and eccentricity. The results are presented
in Fig.~\ref{Fig:IP_phys} and show the absence of any obvious relation
between the initial and final eccentricities. In particular, a system
initially more eccentric will not necessarily end up with a higher final
eccentricity and the reason is because quantities are averaged over an
orbit. For a given initial period, with increasing eccentricity, the stars
get closer to each other at periastron. On one hand, this contributes to
further enhance the mass-loss rate due to its strong dependence on the
Roche radius (which depends on the instantaneous separation) and hence
$\dot{e}_\mathrm{winds}$ but on the other hand, the star spends a longer
fraction of its time at greater distances, where the wind is weak. The
result is that the mean wind mass-loss rate decreases with increasing
eccentricity, leading to less efficient eccentricity pumping and in the end
to a stronger circularization of the orbit. Due to the non-linearity of these
effects, this conclusion may, however, depend on the initial configuration.

\begin{figure}
\includegraphics[width=9cm]{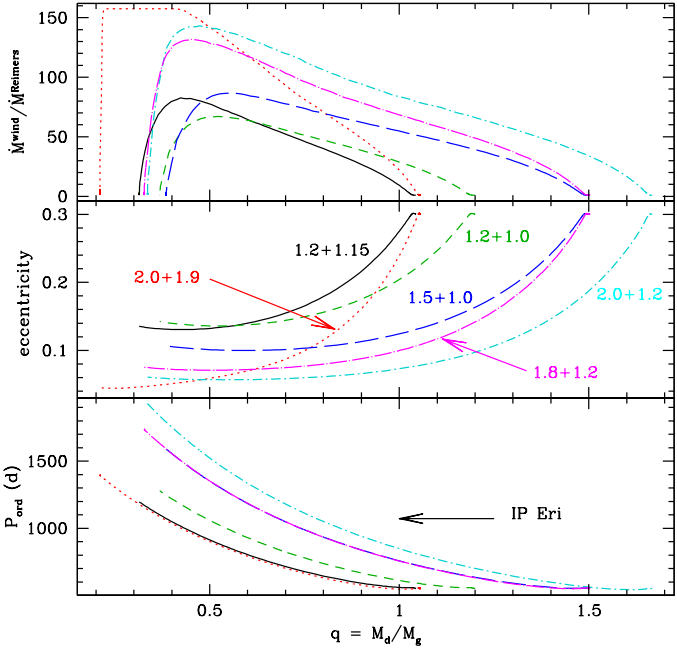}
\caption[]{\label{Fig:mass_ratio} Evolution of the wind-enhancement factor
  (top), eccentricity (mid) and period (bottom) as a function of the mass
  ratio for the different initial configurations specified in the
  graph. All systems have an initial period of 550~d and an eccentricity
  $e=0.3$. The arrow indicates the current orbital period of IP Eri. Note
  the saturation of the mass-loss rate in the 2.0+1.9~\Msun{} system (top panel).}
\end{figure}

Given the uncertainties in the initial-system mass ratio, we performed
several simulations, starting with the same initial period of 550 days and
varying $q$. All mass-losing stars end up with about the same WD mass
between 0.37 and 0.43~\Msun. From the curves depicted
in Fig.~\ref{Fig:mass_ratio}, we see that increasing the donor's mass for a
given companion mass leads to longer final period (green \vs blue curve)
because in this non-conservative evolution, a larger amount of mass has been removed from the system to form the He WD (this
response of the orbital parameters to systemic mass loss is demonstrated in
the appendix). If we now fix the mass of the WD progenitor but increase
that of the companion (cyan \vs red curve), the systems evolve towards
shorter final periods. In these configurations about the same mass is
ejected from the system but, in the binaries with the higher mass companion, a much larger amount of AM will be taken away
by the wind because the specific AM of the ejected material
(Eq.~\ref{eq:jdot}) increases with increasing system mass ($\omega \propto
\sqrt{M_d+M_g}$) and decreasing mass ratio.

Finally, systems with similar mass ratios (red and black curves) evolve
along the same $q$ \vs $P_\mathrm{orb}$ trajectory as demonstrated in the
appendix (Eq.~\ref{eq:period}). This is a consequence of our prescription
for the systemic AM loss rate (Eq.~\ref{eq:jdot}) and of the fact that the
gainer is not accreting mass. As discussed previously, the final periods
lengthen with increasing donor mass.

The final eccentricity (which in Fig.~\ref{Fig:mass_ratio} always ends up
being too small with respect to IP~Eri observed value) is dictated by the
competition between $\dot{e}_\mathrm{winds}$ and $\dot{e}_\mathrm{tides}$
and, for a given initial period, depends mostly on the donor's initial mass
because the $\dot{e}$ contributions are imposed by the mass-losing star
($\dot{e}_{\mathrm{tide},d} \gg \dot{e}_{\mathrm{tide},g}$ and
$\dot{M}^\mathrm{wind}_d$ responsible for $\dot{e}_{\mathrm{winds}}$). We also notice
that systems with higher initial masses require larger initial separations
to avoid RLOF. Consequently, they end their evolution at much longer
periods that are incompatible with observations.

\begin{figure}
\includegraphics[width=9cm]{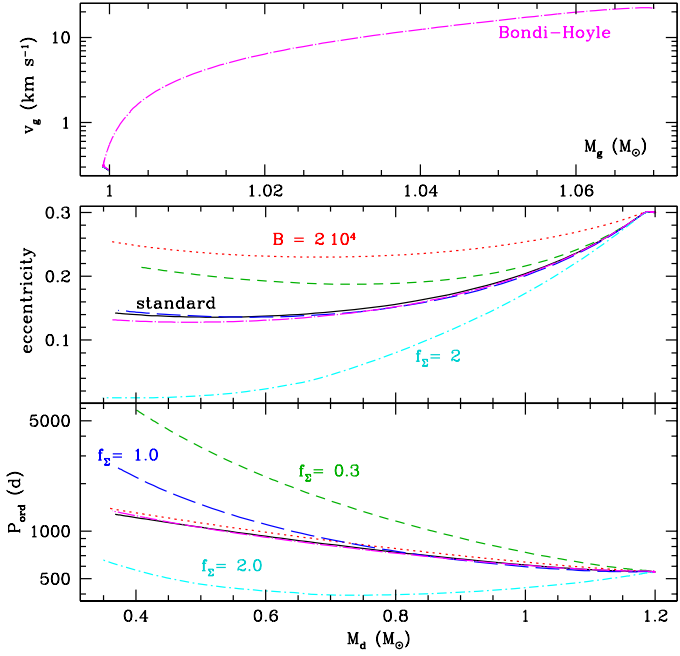}
\caption[]{\label{Fig:rot} Evolution of the surface spin velocity of the
  gainer as a function of its mass (top, with $f_\mathrm{jacc} =0.1$) and
  of the eccentricity (mid) and period (bottom) as a function of the
  donor's mass under different physical assumptions as specified in the
  graph. The initial configuration is a 1.2+1.0 \Msun{}  system, with an
  initial period of 550~d and an eccentricity $e=0.3$ (see text for
  details).}
\end{figure}

Among the various parameters influencing the final orbital period, the
amount of AM lost by the wind is an important one. By default we assume
that the wind carries away the orbital angular momentum of the star at its
position on the orbit (Eq.~\ref{eq:jdot}). In an alternative set of model
calculations depicted in Fig.~\ref{Fig:rot}, we use the prescription
described by Eq.~(37) of \cite{Siess2013}:
\begin{equation}
  \dot{J}_\Sigma = f_\Sigma\, (\dot{M}_{g}^{\mathrm{wind}}+
  \dot{M}_{d}^{\mathrm{wind}}) j_\mathrm{orb}
\end{equation}
where $f_\Sigma$ is a free parameter encapsulating our ignorance of the
exact mode of AM ejection. With decreasing values of this parameter, the
wind removes less AM from the system.  $J_\mathrm{orb}$ is consequently
larger, the separation larger and the tidal torques weaker. The system is
thus able to keep a higher eccentricity and the separation keeps increasing
as a result of non-conservative evolution, but to an extent that is no
longer compatible with the observations (see for example the cases with
$f_\Sigma =$ 0.3 or 1 in Fig.~\ref{Fig:rot}). If instead we set $f_\Sigma=
2.0$, the period remains within the observed value but because of the
shorter separation, the eccentricity is substantially reduced. So this
alternative prescription for $\dot{J}_\Sigma$ cannot at the same time
increase the eccentricity and maintain the period close to $\sim 10^3$~d.

So far, we have neglected the possibility that a fraction of the
tidally-enhanced wind could be captured by the companion. In a first
simulation, we used the classical Bondi-Hoyle wind-accretion scheme where
the default wind parameters of Eq.~(25) of Paper~I are adapted to the slow
($\la 20$~\kms) wind of the giant ($\beta_w = \nicefrac{1}{80}$ and
$\alpha_{\mathrm{BH}} = 0.15$ instead of $\nicefrac{1}{8}$ and 1.5
respectively as suggested by \cite{Hurley2002} for classical Bondi-Hoyle
accretion rates). Following \cite{Shapiro1976} and
\cite{Jeffries-Stevens-96}, we approximate the torque exerted by the wind
onto the gainer by
\begin{equation}
\label{Eq:alternateJ}
  \dot{J}_{\mathrm{acc},g}^\mathrm{wind} =  f_\mathrm{jacc} \frac{ \dot M_\mathrm{acc}^\mathrm{wind} r_\mathrm{acc}^2\omega}{2}
\end{equation}
where $f_\mathrm{jacc}$ is a free parameter set to 0.1 as suggested by
\cite{Jeffries-Stevens-96}. The gainer's accretion radius is given by
\begin{equation}
r_\mathrm{acc} = \frac{2{\cal{G}}M_g}{v^2_\mathrm{orb}+v^{2}_w}
\end{equation}
where $v^2_\mathrm{orb}={\cal{G}}(M_d+M_g)/a$ is the orbital velocity and $v_w$ the wind
velocity set to a fraction $\beta_w$ of the star's escape velocity.

Accounting for wind accretion does not significantly alter the global
picture because little AM is deposited on the gainer. The main observable
effect is an increase of the companion's rotation rate. For reasonable
values of the parameters as used in this test simulation, the gainer
accelerates up to 20~\kms, which is not so far off the observed value. It
is important to emphasize that the enhanced wind scenario avoids the
problem associated with the critical rotation of the gainer star when mass
is transferred via RLOF \citep{Packet81,Dervi2010,Deschamps2013}.

As a final test, we also varied the wind parameter $B_\mathrm{wind}$. This
parameter is badly constrained and is likely to depend on the structure of
the star and vary with time. Considering these large uncertainties
(including the {\it ad hoc} dependence of Eq.~(\ref{eq:mloss}) on
$R/R_{L1}$), we calculated a new model using the standard physics
(i.e. $\dot{J}_\Sigma$ given by Eq.~(\ref{eq:jdot}) and no wind accretion)
but with $B_\mathrm{wind} = 2\times 10^4$ (Fig.~\ref{Fig:rot}, red
curve). As expected, because of the stronger wind, the eccentricity-pumping
mechanism is more efficient leading to a final eccentricity of $\sim 0.23$
in very good agreement with the observed value for IP Eri. This doubling of
the wind mass-loss rate has little impact on the orbital period which is
very similar to our default case.

Finally, Fig.~\ref{Fig:summary} depicts the temporal evolution of some key
observable parameters for the model calculation (initial masses of
1.5+1.45~\Msun, initial period of 415~d, initial eccentricity of 0.4,
$B_{\rm wind} = 3.6\times 10^4$, $\alpha_{\mathrm{BH}}=0.1$, $\beta_w =
\nicefrac{1}{80}$, $f_\mathrm{jacc} = 0.03$ and Eq.~(\ref{eq:jdot}) for
$\dot{J}_\Sigma$) best reproducing the current values of the system
IP~Eri. The agreement is remarkable considering that we are able to fit at
once 7 observational constraints. We note however that the mass of the He
WD (0.35~\Msun) is slightly below the value inferred from model atmosphere
fitting of the WD spectrum. A better agreement could in principle be
achieved if some core overshooting is operating in the giant but one should
also be aware that the procedure used to determine the WD mass suffers its
own limitations and uncertainties. Finally, the slow rotation of the K0
giant can only be fitted if $f_\mathrm{jacc} = 0.03$ implying that little
AM must be accreted.

\section{Discussion}
\label{Sect:conclusion}

The formation of a He WD requires a binary system in which the evolution of
one of the components is truncated before it reaches the tip of
the RGB.  Three main evolutionary channels can lead to a binary system
hosting a He WD and a main sequence or giant star:
\begin{enumerate}
\item unstable RLOF mass transfer. If the initial mass ratio exceeds
  $q_\mathrm{crit} \ga 1.2-1.5$, the mass transfer is dynamical and the system
  most likely enters a common-envelope phase where the eccentricity is
  reduced to zero and the period is considerably shortened;
\item RLOF mass transfer is stable because initially $q < q_\mathrm{crit}$. The
  system avoids a catastrophic evolution and ends up as a long-period binary
  in a circular orbit;
\item the initially most massive star ejected its envelope due to enhanced
  stellar winds and RLOF is avoided. If the period is long enough,
  eccentricity is preserved or can even be increased.
\end{enumerate}

The properties of IP Eri, a system consisting of a giant K0 and a He WD
with a period of 1071~d and an eccentricity of 0.23, can only be explained
by the tidally-enhanced wind model, i.e. via scenario 3 mentioned
above. Our exploration of this evolutionary channel indicates that there is
a critical period below which eccentricity decrease due to tidal effects
will always dominate over the eccentricity pumping due to the wind mass
loss. But such a limit is difficult to determine without an extensive
exploration of the parameter space. As presented in Sect.~\ref{sect:obs},
this scenario also imposes some constraints on the initial mass ratio of IP
Eri. It must not be too far above unity so that in the relatively short
time interval during which the envelope is stripped and the WD cooled down,
the initially lower-mass companion star has left the main sequence and
started to climb the RGB. We also showed that (1) only stars less massive
than 3~\Msun{} may give birth to a He WD (see Sect.~\ref{sect:obs}) and (2)
at solar metallicity, $M_d \ga 1.2\,\msun$ for the He WD to form within
5~Gyr.

\begin{figure}
\includegraphics[width=9cm]{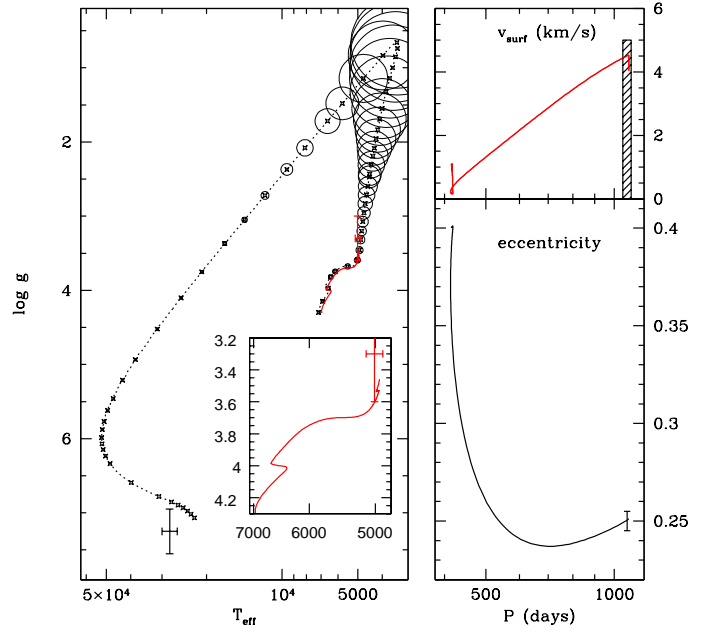}
\caption[]{\label{Fig:summary} Temporal evolution of some key observable
  parameters for the model calculation (initial masses of 1.5+1.45~\Msun,
  initial period of 415~d, $e=0.4$, $B_{\rm wind} = 3.6\times 10^4$,
  $\alpha_{BH}=0.1$, $\beta_w = \nicefrac{1}{80}$ and $f_\mathrm{jacc} =
  0.03$) best reproducing the current values of the system IP~Eri: the left
  panel corresponds to gravity ($\log g$)  \vs $T_{\rm eff}$  (the WD is
  depicted by the black cross, and the K0 star by the red cross, including
  the inset). The circles are proportional to the radius of the mass-losing
  star. The lower right panel is eccentricity versus orbital period, and
  the upper right is spin velocity of the giant versus orbital period. The
  hatched region corresponds to the possible values according to
  observations, which only provide an upper limit on the spin velocity.  }
\end{figure}

In contrast to the related Wind-Induced Rapidly-Rotating (WIRRING) systems
like 2RE~J0357+283 \citep[][]{Jeffries-Stevens-96}, the slow angular
velocity of the K0 giant in IP~Eri indicates that little angular momentum
has been accreted in the He-WD formation process ($f_\mathrm{jacc} =
0.03$). The main reason for this difference is that in their model,
\cite{Jeffries-Stevens-96} consider the ejection of a massive AGB envelope
and impose a significantly higher wind mass-loss rate ($>
10^{-5}$~\myr). They also restrict their study to circular systems, assume
a constant AGB wind mass-loss rate and do not follow the evolution of the
structure of the stellar components which for our RGB is
significant. Indeed, during the tidally-enhanced mass-loss phase, the
radius of the giant increases by more than a factor 4 (see
Fig.~\ref{Fig:summary}), reaching up to 40-60~$R_\odot$ before leaving the
RGB. All these effects result in larger amounts of mass and AM being
accreted by the companion of the AGB star which is spun up to much higher
rotational velocities.

The tidally-enhanced wind model implies that a substantial amount of mass
must be removed from the system. This mass may give rise to a large IR
excess. However, by the time the He WD is observed (some $10^7$ yr after the
end of the envelope ejection phase for our 1.2+1.0~\Msun{} system), this material
has long since been dispersed and mixed with the interstellar medium,
making its detection difficult.

Scenarios 1 and 2 above most likely produce circular systems because of tidal
effects acting before and during RLOF, unless some eccentricity can be
generated either via the interaction between a circumbinary disc and the
orbit \citep{Goldreich1979,Artimowicz1994,Dermine2012} or because of
asymmetric mass loss \citep{Soker2000,Frankowski2001}.

The detection of He-WD + MS or giant binaries, and more recently of sdBs,
with long periods and high eccentricities, support the tidally-enhanced
wind model which is otherwise often advocated to explain the evolution of a
handful of additional systems, including some RS CVn binaries, Algols and
Barium stars as reviewed by \cite{Eggl89}. However the source of
eccentricity may differ between these systems. For example the presence of
a circumbinary disc \citep{Goldreich1979,Artimowicz1994,Dermine2012} could
act in place or in parallel to the asymmetric mass loss mechanism.

Clearly this tidally-enhanced wind model provides a simple
framework to reproduce the observed properties of IP Eri but such studies
need to be extended to different classes of objects, in order to better
constrain the parameters of our models and understand the source of this
puzzling eccentricity.

\begin{acknowledgement}
  This work has been partly funded by an {\it Action de recherche
    concertée} (ARC) from the {\it Direction g\'en\'erale de l'Enseignement
    non obligatoire et de la Recherche scientifique -- Direction de la
    recherche scientifique -- Communaut\'e fran\c{c}aise de Belgique}. LS
  is research associate at the F.R.S.-FNRS. P.J.D. is Senior Research Assistant at F.R.S.-FNRS.
  
\end{acknowledgement}

\bibliographystyle{aa}
\bibliography{siess}

\appendix

\section{Evolution of the period as function of the mass ratio}
In the framework of the tidally-enhanced stellar wind model, the companion
star accretes almost no matter and the rate of change of the orbital AM is
mainly due to the systemic AM loss rate resulting from the giant's wind.
Under these circumstances, $ \dot{J}_{\Sigma} \approx
\dot{J}_{\mathrm{orb}} $ and Eq.~(\ref{eq:adot}) simplifies to
\begin{equation}
  \frac{\dot{a}}{a}\approx 2\frac{\dot{J}_\Sigma}{J_{\mathrm{orb}}}-
  2\frac{\dot{M}_{d}}{M_{d}}+ \frac{\dot{M}_{d}}{M}\ .
  \label{eq:adot2}
\end{equation}
With $q=M_d/M_g$ and $\dot M_g = 0$,  Eq.~(\ref{eq:jdot}) can be recast into 
\begin{equation}
  \dot{J}_\Sigma =  \frac{\dot M^\mathrm{wind}_d}{q}j_{\mathrm{orb}}
  \label{eq:jsigma}
\end{equation}
Substituting Eq.~(\ref{eq:jsigma}) into Eq.~(\ref{eq:adot2}) leads to 
\begin{equation}
  \frac{\dot{a}}{a} = -\frac{{\dot M}_d}{M}\ .
  \label{eq:adot3}
\end{equation}
This last equation tells us that when systemic mass loss prevails and has
the usual form of Eq.~(\ref{eq:jdot}),
the orbital separation increases independently of the mass ratio. Now using
Kepler's third law along with Eq.~(\ref{eq:adot3}), we can relate the change
in period $P$ to that of the system's mass
\begin{equation}
  \frac{d\ln P}{dt} = \frac{3}{2}\frac{d\ln a}{dt} -\frac{1}{2}\frac{d \ln
    M}{dt} = -2 \frac{{\dot M}_d}{M} = -2 \frac{d \ln M}{dt}
  \label{eq:dlnP}
\end{equation}
and with $M=(1+q)M_g$, one finally obtains the relation 
\begin{equation}
d \ln P = -2 \,d \ln(1+q)
\end{equation}
which, after integration leads to
\begin{equation}
 P = P_0 \, \Bigg(\frac{1+q}{1+q_0}\Bigg)^{-2}
\label{eq:period}
\end{equation}
where $P_0$ and $q_0$ are the initial period and mass ratio. This relation
explains why, when systemic mass loss dominates, the evolution of the
period only depends on that of the mass ratio.

\end{document}